\documentclass[twocolumn,showpacs,amsmath,amssymb,10pt,aps]{revtex4}
\usepackage{graphicx,color}
\usepackage{amsmath}
\usepackage{amssymb}
\usepackage{bbm}
\usepackage{color}

\usepackage[utf8]{inputenc}

\usepackage{amsfonts}
\usepackage{bm}

\newcommand{\Md}{M_d(\mathbb{C})}

\newcommand{\ra}{{\rightarrow}}
\newcommand{{\Cd}}{{\mathbb{C}^d}}

\newcommand{\bra}[1]{\mbox{$\langle #1 |$}}

\newcommand{\ket}[1]{\mbox{$| #1 \rangle$}}
\def\oper{{\mathchoice{\rm 1\mskip-4mu l}{\rm 1\mskip-4mu l}
{\rm 1\mskip-4.5mu l}{\rm 1\mskip-5mu l}}}
\def\<{\langle}
\def\>{\rangle}

\newtheorem{Remark}{Remark}

\newcommand{\beq}{\begin{equation}}
\newcommand{\eeq}{\end{equation}}
\newcommand{\bear}{\begin{eqnarray}}
\newcommand{\ear}{\end{eqnarray}}
\newcommand{\bdm}{\begin{displaymath}}
\newcommand{\edm}{\end{displaymath}}

\begin{document}
\title{\textbf{
Generalized Pauli channels and a class of non-Markovian quantum evolution}}
\author{Dariusz Chru\'sci\'nski and Katarzyna Siudzi\'nska}
\affiliation{ Institute of Physics, Faculty of Physics, Astronomy and Informatics \\  Nicolaus Copernicus University,
Grudzi\c{a}dzka 5/7, 87--100 Toru\'n, Poland}



\begin{abstract}
We analyze the quantum evolution represented by a time-dependent family of generalized Pauli channels. This evolution is provided by the random decoherence channels with respect to the maximal number of mutually unbiased bases. We derive the necessary and sufficient conditions for the vanishing back-flow of information.

\end{abstract}

\maketitle

\section{Introduction}

Recently, much effort was devoted to analyze the non-Markovian evolution of open quantum systems \cite{Book,Weiss}. The current state of the art is summarized in recent review papers  \cite{rev1,rev2}.

Let us recall that an arbitrary quantum evolution may be represented by a dynamical map, that is,  a family of completely positive, trace-preserving maps $\Lambda(t) : \mathcal{B}(H) \rightarrow \mathcal{B}(\mathcal{H})$ ($t \geq 0$) satisfying the initial condition $\Lambda(0) = \oper$ (the identity map). $\Lambda(t)$ maps the initial state represented by the density operator $\rho$ into the current state $\rho(t)  = \Lambda(t)[\rho]$. It is therefore clear that the Markovianity property of quantum evolution is controlled by certain properties of the corresponding dynamical map $\Lambda(t)$. The two most popular  approaches  are based on divisibility of the corresponding dynamical map \cite{RHP,Wolf-Isert} and distinguishability of states  \cite{BLP}. The map $\Lambda(t)$ is CP-divisible iff
\begin{equation}\label{CP-div}
  \Lambda(t) = V(t,s) \Lambda(s)
\end{equation}
and $V(t,s)$ is CPTP for all $t \geq s$. If $V(t,s)$ is only positive, $\Lambda(t)$ is called P-divisible --- this property is related to the Markovianity of classical stochastic evolution \cite{P-DIV} (actually, one may introduce the whole hierarchy of $k$-divisible maps, for which $V(t,s)$ is a $k$-positive map \cite{PRL-Sabrina}). Following \cite{RHP}, one calls a quantum evolution Markovian iff $\Lambda(t)$ is CP-divisible. Interestingly, this property is fully controlled by the corresponding time-local generator $\mathcal{L}(t)$ entering the time-local master equation
\begin{equation}
  \frac{d}{dt} \Lambda(t) =\mathcal{L}(t) \Lambda(t) .
\end{equation}
Namely, $\Lambda(t)$ is CP-divisible if and only if $\mathcal{L}(t)$ has the standard time-dependent GKSL \cite{GKSL} form, $  \mathcal{L}(t)[\rho] = -i[H(t),\rho] + \mathcal{L}_D(t)[\rho]$,
with the dissipative part
\begin{equation*}
\mathcal{L}_D(t)[\rho] = \sum_\mu \gamma_\mu(t)( [V_\mu(t),\rho V_\mu^\dagger(t)] +    [V_\mu(t)\rho, V_\mu^\dagger(t)]) ,
\end{equation*}
and $\gamma_\mu(t) \geq 0$. In this case, $\gamma_\mu(t)$ may be interpreted as time-dependent decoherence/dissipation rates.  A slightly weaker condition for Markovianity was proposed in \cite{BLP}: $\Lambda(t)$ represents the Markovian evolution iff
\begin{equation}\label{BLP}
  \frac{d}{dt} ||\Lambda(t)[\rho_1-\rho_2]||_1 \leq 0
\end{equation}
for any pair of initial states $\rho_1$ and $\rho_2$ ($||X||_1 = {\rm Tr} \sqrt{XX^\dagger}$ denotes the trace norm of $X$). The above condition is interpreted as no back-flow of information (i.e. no information {\em flows} from the environment into the system enabling one to increase distinguishability of states of the system).  The CP-divisibility implies (\ref{BLP}), but the converse is not true \cite{Angel} (see also recent studies in \cite{Buscemi,Toni}). Condition (\ref{BLP}) is a special case of a more restrictive condition:
\begin{equation}\label{P-div}
  \frac{d}{dt} ||\Lambda(t)[X]||_1 \leq 0
\end{equation}
for all Hermitian $X$ ((\ref{BLP}) takes into account only traceless Hermitian operators). It turns out that, as long as $\Lambda(t)$ is invertible, condition (\ref{P-div}) is equivalent to  P-divisibility \cite{PRL-Sabrina}. The most appealing feature of condition (\ref{BLP}) is its operational meaning, since $||\rho_1 - \rho_2||_1$ measures the distinguishability of the states $\rho_1$ and $\rho_2$. However, contrary to the CP-divisibility, condition (\ref{BLP}), in general, does not have a natural interpretation on the level of the time-local generator $\mathcal{L}(t)$.


In this paper, we analyze a special class of dynamical maps $\Lambda(t)$  --- the generalized Pauli channels --- and provide the criteria for its Markovian/non-Markovian behaviour. The Pauli channels and their generalizations \cite{Ruskai,Petz} proved to be important in many aspects of quantum information theory \cite{QIT}.
The crucial ingredient in their construction is the set of mutually unbiased bases (MUBs) \cite{Wootters} (see \cite{MUB-Karol} for the review). Again, there are already many important applications of MUBs in quantum tomography \cite{Wootters,T0,T1}, quantum cryptography \cite{Cr1,Cr2},  the mean king's problem \cite{MK1,MK2}, and entropic uncertainty relations \cite{Entr1,Entr2,Entr3}. Recently, they were used to witness entangled quantum states \cite{Beatrix,W-MUB}.

We consider the quantum evolution provided by the random decoherence channels with respect to the maximal number of MUBs and derive necessary and sufficient conditions for no back-flow of information in terms of the corresponding time-local generator.


\section{Weyl channels and generalized Pauli channels}

The celebrated Pauli channel $\Lambda_P : M_2(\mathbb{C}) \ra  M_2(\mathbb{C})$,
\begin{equation}\label{Pauli}
  \Lambda_P[\rho] = \sum_{\alpha=0}^3 p_\alpha \sigma_\alpha \rho \sigma_\alpha ,
\end{equation}
may be easily generalized to the Weyl channel $\Lambda_W : M_d(\mathbb{C}) \ra  M_d(\mathbb{C})$,
\begin{equation}\label{W}
  \Lambda_W[\rho] = \sum_{k,l=0}^{d-1} p_{kl} W_{kl} \rho W_{kl}^\dagger ,
\end{equation}
where $W_{kl}$ are the unitary Weyl operators defined by
$$
W_{kl}=\sum_{m=0}^{d-1} \omega^{mk}\ket{m}\bra{m+ l},
$$
with $\omega=e^{2\pi i/d}$. They satisfy the well-known relations:
$$
W_{kl}W_{rs}=\omega^{ks}W_{k + r,l + s},\quad W_{kl}^\dag=\omega^{kl}W_{-k,-l}.
$$
Clearly, for $d=2$ the Weyl channel simplifies to the Pauli channel.

Note  however, that passing  from (\ref{Pauli}) to (\ref{W}), one loses the most characteristic features of the Pauli channel. The original Pauli channel is Hermitian, that is, ${\rm Tr}(A \Lambda[B]) =  {\rm Tr}(\Lambda[A]B)$. Hence, it possesses real eigenvalues $\{\lambda_0=1,\lambda_1,\lambda_2,\lambda_3\}$ related to the probability vector $p_\alpha$ via
\begin{equation}
  \lambda_\alpha = p_0 + p_\alpha - \sum_{\beta \neq \alpha} p_\beta
\end{equation}
for $\alpha=1,2,3$. The inverse relation reads
\begin{equation}
  p_0 = \frac 14(1 + \lambda_1 + \lambda_2 + \lambda_3)
\end{equation}
and
\begin{equation}
  p_\alpha = \frac 14 \left( 1 + \lambda_\alpha - \sum_{\beta \neq \alpha} \lambda_\beta \right)
\end{equation}
for $\alpha=1,2,3$.
Actually, this relation allows one to derive the well-known Fujiwara-Algoet condition \cite{Fujiwara,King,Szarek,Karol},
\begin{equation}\label{F-A}
  |1\pm \lambda_3| \geq |\lambda_1 \pm \lambda_2| ,
\end{equation}
which guarantees that (\ref{Pauli}) defines a CP map. Note that the above condition may be rewritten as follows,
\begin{equation}\label{Fuji-2}
  -  1 \leq \sum_{\beta =1}^{3} \lambda_\beta \leq 1 + 2 \min_\beta \lambda_\beta .
\end{equation}
The Weyl channel (\ref{W}) is no longer Hermitian for $d>2$, and hence the corresponding eigenvalues are in general complex \cite{Filip-PRA}. The second important feature of the Pauli channel is related to the well-known fact that the eigenbases of $\{\sigma_1,\sigma_2,\sigma_3\}$ are mutually unbiased. Let us recall that two orthonormal bases $|\psi_k\>$ and $|\phi_l\>$ in $\Cd$ define mutually unbiased bases (MUBs) iff for any $k$ and $l$ the following condition is satisfied:
\begin{equation}
  |\<\psi_k|\phi_l\>|^2 = \frac 1 d .
\end{equation}
Moreover, it is well-known \cite{Wootters} that the number $N(d)$ of MUBs in $\mathbb{C}^d$ is bounded by $N(d) \leq d+1$ \cite{MAX}. If $d=p^r$ with $p$ being a prime number, one has $N(d)=d+1$. In this case, explicit constructions are known \cite{Wootters,MAX}. If $d=d_1d_2$, then $N(d) \geq \min \{N(d_1),N(d_2)\}$ \cite{MUB-1}. Moreover, Grassl \cite{MUB-2} provided a construction of three MUBs in an arbitrary dimension.

Let us assume that the $d$-dimensional  Hilbert space admits the maximal number $N(d)=d+1$  of MUBs,
$  \{ |\psi^{(\alpha)}_0\>,\ldots,|\psi^{(\alpha)}_{d-1}\> \}$, and denote the corresponding rank-1 projectors by $P^{(\alpha)}_l = |\psi^{(\alpha)}_l\>\< \psi^{(\alpha)}_l|$. Now, define $d+1$ unitary operators
\begin{equation}\label{U}
  U_{\alpha} = \sum_{l=0}^{d-1} \omega^{l} P_l^{(\alpha)} \ ,
\end{equation}
with $\omega = e^{2\pi i/d}$, and $d+1$ CP maps
\begin{equation}
  \mathbb{U}_\alpha[\rho] = \sum_{k=1}^{d-1}  U_{\alpha}^k \rho  U_{\alpha}^{k \dagger} .
\end{equation}
Note that
\begin{equation}\label{U-Phi}
   \mathbb{U}_\alpha = d \Phi_\alpha - \oper ,
\end{equation}
where
\begin{equation}
  \Phi_\alpha[\rho] =  \sum_{l=0}^{d-1}  P_l^{(\alpha)} \rho  P_l^{(\alpha)} ,
\end{equation}
is the quantum-classical channel. Channels $\Phi_\alpha$  satisfy the following properties:
\begin{equation}
  \Phi_\alpha \Phi_\alpha = \Phi_\alpha ,
\end{equation}
and for $\alpha \neq \beta$ we have
\begin{equation}\label{Phi-Phi}
   \Phi_\alpha \Phi_\beta[\rho] =  \Phi_\beta \Phi_\alpha[\rho] = \frac 1d \mathbb{I} {\rm Tr}\rho .
\end{equation}
The consecutive action of two different channels, $\Phi_\alpha$ and $\Phi_\beta$, gives rise to the complete depolarization of $\rho$. Therefore, it is clear that
\begin{equation}\label{Phi-Phi-Phi}
  \Phi_{\alpha_1} \ldots  \Phi_{\alpha_k}[\rho] =   \frac 1d \mathbb{I} {\rm Tr}\rho
\end{equation}
if there are at least two different $\alpha$'s in the set $\{ \alpha_1 , \ldots  , \alpha_k \}$.
Moreover,
\begin{equation}
  \Phi_\alpha[U_\beta^k] = \delta_{\alpha\beta} U_\beta^k \ ; \ k=1,\ldots,d-1 .
\end{equation}
For $d=2$ one has three MUBs:
\begin{eqnarray*}
  \mathcal{B}_1 &=& \left\{ \frac{|0\> + |1\>}{\sqrt{2}},  \frac{|0\> - |1\>}{\sqrt{2}} \right\} , \\
   \mathcal{B}_2 &=& \left\{ \frac{|0\> + i|1\>}{\sqrt{2}},  \frac{|0\> - i|1\>}{\sqrt{2}} \right\} ,  \\
   \mathcal{B}_3 &=& \{ |0\>,|1\>\} ,
\end{eqnarray*}
and $\mathbb{U}_\alpha[\rho] = \sigma_\alpha \rho \sigma_\alpha$ for $\alpha=1,2,3$.
Finally, let us define the generalized Pauli channel,
\begin{equation}
  \Lambda_{\rm GP} = p_0 \oper + \frac{1}{d-1}\sum_{\alpha=1}^{d+1} p_\alpha \mathbb{U}_\alpha ,
\end{equation}
where $(p_0,p_1,\ldots,p_{d+1})$ is the probability vector. Such channels were already analyzed in \cite{Ruskai} under the name {\em Pauli channels constant on axes}. The crucial property of $\Lambda_{\rm GP}$ consists in
\begin{equation}
   \Lambda_{\rm GP}[ U_\alpha^k] = \lambda_\alpha  U_\alpha^k  \ , \ \ k=1,\ldots,d-1 ,
\end{equation}
where
\begin{equation}
  \lambda_\alpha = p_0 + p_\alpha - \frac{1}{d-1} \sum_{\beta \neq \alpha} p_\beta ,
\end{equation}
and $\lambda_0=1$ due to the fact that $\Lambda_{\rm GP}$ is unital, i.e. $ \Lambda_{\rm GP}[\mathbb{I}]=\mathbb{I}$. All the eigenvalues are real and $\lambda_\alpha$ are $(d-1)$ times degenerated (apart from $\Lambda_0$). The inverse relation reads:
\begin{equation}\label{c1}
  p_0 = \frac{1}{d^2}(1 + (d-1)[\lambda_1 + \ldots + \lambda_{d+1}]) ,
\end{equation}
and
\begin{equation}\label{c2}
  p_\alpha = \frac{d-1}{d^2} \left( 1 + (d-1)\lambda_\alpha -  \sum_{\beta \neq \alpha} \lambda_\beta \right) ,
\end{equation}
for $\alpha =1,\ldots,d+1$. From (\ref{c1})-(\ref{c2}), it is clear that $\Lambda_{\rm GP}$ is CP if and only if \cite{Ruskai}
\begin{equation}\label{Fuji-d}
  -\frac{1}{d-1} \leq  \sum_{\beta =1}^{d+1} \lambda_\beta \leq 1 + d \min_\beta \lambda_\beta ,
\end{equation}
which is a direct generalization of (\ref{Fuji-2}).

\begin{Remark} Actually, for prime dimensions one may construct a complete set of MUBs using the Weyl operators. Fixing the orthonormal basis $\{|0\>,|1\>,\ldots,|d-1\>\}$ in $\mathcal{H}$, one introduces $X$ and $Z$ by
\begin{equation}
  X|k\>=|k+1\> , \ \ Z|k\> = \omega^k|k\> ,
\end{equation}
with $\omega= e^{2\pi i /d}$. The corresponding Wyel operators read as follows,
\[W_{kl}=X^lZ^k.\]
If $d$ is a prime number, then the following set of unitary operators,
\[Z,\ X,\ XZ,\ \dots,\ XZ^{d-1},\]
gives rise to $d+1$ MUBs --- these are the eigenbases of $Z$ and $XZ^k$ for $k=1,\ldots,d-1$. Hence, the eigenbases of
$$ W_{01},W_{10},W_{11},\ldots,W_{1,d-1} $$
define $d+1$ MUBs. Moreover, if $d$ is prime, any orthonormal basis $\mathcal{B}=\{U_0=\mathbb{I},U_1,\ldots,U_{d^2-1}\}$ in $M_d(\mathbb{C})$ consisting of unitary matrices splits into the following classes,
\begin{equation}
  \mathcal{B} = \{\mathbb{I}\} \cup \mathcal{C}_1 \cup \ldots \cup \mathcal{C}_{d+1} .
\end{equation}
Each class $\mathcal{C}_k$ contains exactly $d-1$ mutually commuting matrices from $\mathcal{B}$ \cite{MAX}. Let us observe that the following  Weyl operators $W_{\alpha k, \alpha l}$ with $\alpha=1,\ldots,d-1$ belong to the same class of mutually commuting unitary operators. For example, if $d=3$, then one has 4 sets of mutually commuting matrices among $8$ Weyl operators:
$$ \{W_{01},W_{02}\} \cup  \{W_{10},W_{20}\} \cup \{ W_{11},W_{22}\} \cup \{W_{12},W_{21}\} . $$
\end{Remark}

\begin{Remark}

The above construction may be generalized if we replace the commutative subalgebras $\mathcal{C}_i$ by the non-commutative subalgebras $\mathcal{A}_i$ of $M_d(\mathbb{C})$. One calls two subalgebras, $\mathcal{A}_1$ and $\mathcal{A}_2$, complementary iff  $\mathcal{A}_1\ominus \mathbb{C} \mathbb{I}$ and $\mathcal{A}_2 \ominus \mathbb{C}\mathbb{I}$ are mutually orthogonal with respect to the Hilbert-Schmidt inner-product \cite{Petz}. Now, suppose that
$$ M_d(\mathbb{C}) = \mathbb{C}\mathbb{I} \cup \mathcal{A}_1 \cup \ldots \cup \mathcal{A}_r  , $$
and $\mathcal{A}_i$ are mutually complementary. Let $\Phi : M_d(\mathbb{C}) \rightarrow \mathcal{A}_i$ denote the corresponding CPTP projector. Then, the map
\[\Phi[\rho]=\left(1-\sum_{i=1}^r\lambda_i\right)\frac{\mathbb{I}}{d} \mathrm{Tr}\rho  +\sum_{i=1}^r\lambda_i \Phi_i[\rho],\]
provides a generalization of the Pauli channel. Such generalized Pauli channels were analyzed by Petz and Ohno \cite{Petz}.
In this paper, we consider only the abelian case corresponding to the commutative algebras $\mathcal{C}_i$, which are related to the families of MUBs.
\end{Remark}

\section{Time-local description of generalized Pauli evolution}

Now, consider the following time-dependent generalized Pauli channel,
\begin{equation}\label{LAMBDA}
\Lambda(t) = p_0(t) \oper + \frac{1}{d-1} \sum_{\alpha=1}^{d+1} p_\alpha(t) \mathbb{U}_\alpha ,
\end{equation}
satisfying the time-local master equation
\begin{equation}
  \frac{d}{dt} \Lambda(t) = \mathcal{L}(t) \Lambda(t)
\end{equation}
with $\Lambda(0) = \oper$.  The corresponding time-local generator $\mathcal{L}(t)$ has the following form,
\begin{equation}\label{GEN}
  \mathcal{L}(t) = \sum_{\alpha=1}^N \gamma_\alpha(t) \mathcal{L}_\alpha ,
\end{equation}
where
\begin{equation}   \label{L-alpha}
  \mathcal{L}_\alpha[\rho]= \frac 1d \left( \mathbb{U}_\alpha[\rho] - [d-1]\rho \right) .
\end{equation}
Again, for $d=2$ one can recover the standard generator,
\begin{equation}
  \mathcal{L}(t) = \frac 12 \sum_{\alpha=1}^3 \gamma_\alpha(t) [\sigma_\alpha \rho \sigma_\alpha - \rho] .
\end{equation}
Taking (\ref{U-Phi}) into account, one arrives at the following simple form:
\begin{equation}
  \mathcal{L}_\alpha = \Phi_\alpha - \oper .
\end{equation}
The property shown in equation (\ref{Phi-Phi}) implies that $[\mathcal{L}_\alpha,\mathcal{L}_\beta]=0$, and hence
\begin{equation}
  \Lambda(t) = \prod_{\alpha=1}^N \Lambda_\alpha(t)
\end{equation}
with
\begin{equation}
  \Lambda_\alpha(t) = e^{\Gamma_\alpha(t)  \mathcal{L}_\alpha} = e^{-\Gamma_\alpha(t)} \oper + (1- e^{-\Gamma_\alpha(t)}) \Phi_\alpha
\end{equation}
and $\Gamma_\alpha(t) = \int_0^t \gamma_\alpha(u)du$.
Note that
\begin{equation}
  \mathcal{L}(t)[U^k_\alpha] = \mu_\alpha(t) U^k_\alpha ,
\end{equation}
where $\mu_\alpha(t) = \gamma_\alpha(t) - \gamma(t)$ and $\gamma(t) = \gamma_1(t) + \ldots + \gamma_{d+1}(t)$. The corresponding time-dependent eigenvalues of the dynamical map $\Lambda(t)$ read:
\begin{equation}
  \lambda_\alpha(t) = \exp[\Gamma_\alpha(t) - \Gamma(t)] ,
\end{equation}
$\Gamma(t) = \Gamma_1(t) + \ldots + \Gamma_{d+1}(t)$. Hence, the generator (\ref{GEN}) gives rise to a CPTP dynamical map if and only if
\begin{equation}
e^{\Gamma_1(t)} + \ldots +  e^{\Gamma_{d+1}(t)} \leq e^{\Gamma(t)} + d \min_{\beta} e^{\Gamma_\beta(t)} ,
\end{equation}
which is a direct consequence of (\ref{Fuji-d}) and the fact that $e^{\Gamma_\beta(t)} \geq 0$.

\section{Non-Markovianity criteria}

The evolution $\Lambda(t)$, generated by $\mathcal{L}(t)$ and defined in (\ref{GEN}), is CP-divisible if and only if $\gamma_\alpha(t) \geq 0$ for $\alpha =1,\ldots,d+1$. Let us analyze the conditions for P-divisibility, which are given by
\begin{equation}\label{P1-d}
  \frac{d}{dt} || \Lambda(t)[X]||_1 \leq 0
\end{equation}
for all Hermitian operators $X$. Taking $X = U_\alpha$, one immediately finds
\begin{equation}\label{P-d}
  \gamma(t) - \gamma_\alpha(t) = \sum_{\beta \neq \alpha} \gamma_\beta(t) \geq 0 ,
\end{equation}
for $\alpha =1,\ldots,d+1$. Hence, conditions (\ref{P-d}) are necessary for P-divisibility. It means that the violation of (\ref{P-d}) implies that $\Lambda(t)$ is not P-divisible, and therefore it is also non-Markovian. For $d=2$ these conditions simplify to
\begin{equation}\label{P-2}
  \gamma_1(t) +   \gamma_2(t) \geq 0 ,   \gamma_2(t) +   \gamma_3(t) \geq 0 ,  \gamma_3(t) +   \gamma_1(t) \geq 0  .
\end{equation}
It was shown in \cite{Filip-PLA,Filip-PRA} that they are necessary and sufficient for P-divisibility. However, for $d > 2$ these conditions are no longer sufficient.

Let us observe that condition (\ref{P-d}) implies that
\begin{equation}\label{P2-d}
  \frac{d}{dt} || \Lambda(t)[X]||_2 \leq 0 ,
\end{equation}
for all Hermitian $X$ ($||X||^2_2 = {\rm Tr} X^\dagger X$ denotes the Frobenius norm).  Indeed, if
\begin{equation*}
  X = x_0 \mathbb{I} + \sum_{\alpha=1}^{d+1} \sum_{k=1}^{d-1} x_{\alpha,k} U_\alpha^k ,
\end{equation*}
then
\begin{equation*}
  \Lambda(t)[X] = x_0 \mathbb{I} + \sum_{\alpha=1}^{d+1} \lambda_\alpha(t) \sum_{k=1}^{d-1} x_{\alpha,k} U_\alpha^k .
\end{equation*}
Calculating the Frobenius norm, we get
\begin{eqnarray*}
  ||\Lambda(t)[X]||_2^2 &=& {\rm Tr}( \Lambda(t)[X^\dagger]\Lambda(t)[X] )\\ &=& d \left(|x_0|^2 + \sum_{\alpha=1}^{d+1} \lambda^2_\alpha(t) \sum_{k=1}^{d-1} |x_{\alpha,k}|^2 \right),
\end{eqnarray*}
from which it follows that
\begin{equation}\label{2-2}
  \frac{d}{dt}||\Lambda(t)[X]||_2^2 = 2d\sum_{\alpha=1}^{d+1} \dot{\lambda}_\alpha(t) \lambda_\alpha(t)\sum_{k=1}^{d-1} |x_{\alpha,k}|^2 \leq 0
\end{equation}
due to $\dot{\lambda}_\alpha(t) \leq 0$ and ${\lambda}_\alpha(t) \geq 0$. Hence, conditions (\ref{P-d}) imply (\ref{P2-d}). However, being sufficient for (\ref{P2-d}), conditions (\ref{P-d}) are not sufficient for (\ref{P1-d}). Therefore, they do not guarantee P-divisibility.

The reason why these conditions are not sufficient is related to the well-known properties of the maps which are contractions \cite{Bhatia,Paulsen}.
Let us recall that if $\Phi$ is a trace-preserving map, then $\Phi$ is positive if and only if $||\Phi[X]||_1 \leq ||X||_1$. Similarly, if $\Phi$ is a unital map, then $\Phi$ is positive if and only if $||\Phi[X]|| \leq ||X||$, where $||X||$ is the operator norm (i.e. $||X||$ is the maximal singular value of $X$).
Now, if $\Phi$ is positive, trace-preserving, and unital (bistochastic), then, apart being a contraction in the trace and operator norm, it is also a contraction in the Hilbert-Schmidt (or Frobenius) norm $||\Phi[X]||_2 \leq ||X||_2$ for all normal $X$. Indeed, using the Kadison inequality,
\begin{equation}
  \Phi[X^\dagger X] \geq \Phi[X^\dagger] \Phi[X] ,
\end{equation}
one has
\begin{eqnarray}
  || \Phi[X]||_2^2 &=& {\rm Tr}(\Phi[X^\dagger] \Phi[X]) \leq {\rm Tr}(\Phi[X^\dagger X]) \nonumber \\ &=& {\rm Tr}(X^\dagger X) = ||X||_2^2 .
\end{eqnarray}
Note that, in general, the above argument cannot be reversed: a trace-preserving unital contraction in the Frobenius norm  does not need to be a positive map.

To derive the sufficient conditions, let us observe that our generator has the following structure:
\begin{equation}
  \mathcal{L}(t) = \frac 1d \left\{ \Phi(t) - 2 \gamma_0(t) \oper \right\} ,
\end{equation}
where
\begin{equation}
  \Phi(t)[\rho] = \sum_{\alpha=1}^{d+1} \gamma_\alpha(t) \sum_{k=1}^{d-1} U_\alpha^k \rho  U_\alpha^{k\dagger} + \gamma_0(t) \rho
\end{equation}
and $\gamma_0(t) = (d-1) \sum_{\alpha=1}^{d+1} \gamma_\alpha(t) $. We make use of the following result \cite{CMP}: let
\begin{equation}
  \Phi[\rho] = \sum_{\mu=1}^{d^2} a_\mu V_\mu \rho V_\mu^\dagger ,
\end{equation}
where $V_\mu$ defines an orthonormal basis of unitary operators in $\Md$. Clearly, if for all $\mu$ we have $a_\mu \geq 0$, then $\Phi$ is CP. Suppose that
$ a_1,\ldots,a_{d-1} \leq 0$. If the remaining $a_\mu$ (for $\mu \geq d$) satisfy
\begin{equation}
  a_\mu \geq |a_1| + \ldots + |a_{d-1}| ,
\end{equation}
then $\Phi$ is a positive map. Now, suppose that only a single  $\gamma_\alpha(t)$ may be strictly negative for $t \geq 0$. Assuming that  $\gamma_{d+1}(t) < 0$, the above result says: if
\begin{equation}\label{suf}
  \gamma_\alpha(t) \geq (d-1)|\gamma_{d+1}(t)| ,
\end{equation}
then $\Phi(t)$ is a positive map, and hence $\Lambda(t)$ is P-divisible. This condition can be rewritten in a more general way:
\begin{equation}\label{SUF}
  \gamma_\alpha(t) + (d-1)\gamma_{\beta}(t) \geq 0
\end{equation}
for any $\alpha \neq \beta = 1, \ldots,d+1$. For $d=2$ the above condition reduces to (\ref{P-2}). Therefore, it provides another proof that (\ref{P-2}) are both necessary and sufficient. In \cite{Filip-PRA} it is shown that for the general Weyl generator
\begin{equation}\label{LW}
  \mathcal{L}_W(t)[\rho] = \sum_{k+l>0}^{d-1} \gamma_{kl}(t)[ W_{kl} \rho W_{kl}^\dagger - \rho] ,
\end{equation}
the sufficient condition for P-divisibility of the corresponding dynamical map $\Lambda_W(t) = \exp( \int_0^t  \mathcal{L}_W(\tau)d\tau)$ reads as follows: assuming that at any time not more than $d-1$ rates $\gamma_\alpha(t)$  are negative, one has
\begin{equation}\label{Wd}
  \gamma_{\alpha_1}(t) + \ldots + \gamma_{\alpha_d}(t) \geq 0 \ ,
\end{equation}
for any $\alpha_1,\ldots,\alpha_d \in \{1,\ldots,d^2-1\}$, where we rearranged the indices according to $(i,j) \rightarrow \alpha = i + dj$. For $d=2$ this
condition again reproduces (\ref{P-2}). Now, in the case of generalized Pauli channel $d^2-1$ rates are grouped into $N(d)=d+1$ sets of identical $d-1$ rates. Hence, only a single rate may be negative. In this case condition (\ref{Wd}) reduces to (\ref{SUF}).


\section{Example --``eternal non-Markovianity''}

In \cite{Erika}, Andersson et. al. provided an example of the qubit evolution generated by
\begin{equation}
  \mathcal{L}(t)[\rho] = \frac 12 \sum_{\alpha=1}^3 \gamma_\alpha(t) [ \sigma_\alpha \rho \sigma_\alpha - \rho] ,
\end{equation}
where $\gamma_1=\gamma_2=1$ and $\gamma_3(t) =-\tanh t$. Note that $\Gamma_3(t) = - \ln[\cosh t] < 0$ for $t>0$. Nevertheless, condition (\ref{Fuji-2}) is satisfied, and hence the corresponding map,
\begin{equation}
  \Lambda(t)[\rho] = \frac{1+ e^{-2t}}{2} \rho + \frac{1 - e^{-2t}}{4}[ \sigma_1 \rho \sigma_1 + \sigma_2 \rho \sigma_2 ] ,
\end{equation}
is completely positive. Due to the fact that $\gamma_3(t) < 0$ for  $t>0$, the authors of \cite{Erika} called this evolution ``eternally non-Markovian''.
Interestingly, being ``eternally non-Markovian'', it satisfies (\ref{P-2}), and therefore it is BLP-Markovian. It turns out \cite{Filip-PRA} that $\Lambda(t)$ is a convex combination of two Markovian semigroups,
\begin{equation}   \label{2}
  \Lambda(t) = \frac 12 \left( e^{2t \mathcal{L}_1} +  e^{2t \mathcal{L}_2} \right)  .
\end{equation}
Consider now (\ref{2}) with $\mathcal{L}_\alpha$ defined in (\ref{L-alpha}). Simple calculations lead to
\begin{equation}\label{}
  \gamma_1(t)=\gamma_2(t) = 1 + \frac{d-2}{d} \tanh t ,
\end{equation}
and
\begin{equation}\label{}
   \gamma_3(t)= \ldots =\gamma_{d+1}(t) = - \frac{2}{d} \tanh t ,
\end{equation}
which proves that among $\{\gamma_1(t),\ldots,\gamma_{d+1}(t)\}$ one may have $d-1$ $\gamma$s which are negative. Note, that in terms of the Weyl channel it means that among $d^2-1$ $\gamma$s one may have $(d-1)^2$ negative elements and still $\Lambda(t)$ is legitimate dynamical map. Note, that conditions (\ref{P-d}) are satisfied. Indeed, one has
$$   \gamma_1(t) + (d-1)\gamma_3(t) = 1 - \tanh t \geq 0 . $$
Consider now the following convex combination of Markovian semigroups:
\begin{eqnarray}
  \Lambda(t) &=& \frac 1d \left( e^{dt \mathcal{L}_1} + \ldots + e^{dt \mathcal{L}_d} \right)  \nonumber \\
  &=& e^{-dt} \oper + (1 - e^{-dt})   \overline{\Phi} ,
\end{eqnarray}
where
\begin{equation}\label{}
  \overline{\Phi} = \frac 1d (\Phi_1 + \ldots + \Phi_d) ,
\end{equation}
that is, $\Lambda(t)$ is a convex combination of the identity map and the `averaged' decoherence channel $\overline{\Phi}$.
Now, using (\ref{U-Phi}) one obtains
\begin{eqnarray}
  \Lambda(t) = \frac 1d \left\{ (1+[d-1]e^{-dt}) \oper + \frac{1-e^{-dt}}{d} (\mathbb{U}_1 + \ldots + \mathbb{U}_d ) \right\} . \nonumber
\end{eqnarray}
One easily shows that the corresponding time-local generator is defined by $\gamma_1= \ldots = \gamma_d = 1$ and
\begin{equation}
  \gamma_{d+1}(t) = -(d-1)\frac{e^{dt} - 1}{e^{dt} + d-1} ,
\end{equation}
which reduces to `$-\tanh t$' for $d=2$. It satisfies the necessary conditions (\ref{P-d}) but violates the sufficient ones (\ref{SUF}).  Numerical analysis shows that it is not P-divisible. This again proves the fundamental difference between the qubit and general qudit cases.

\section{Conclusions}

We analyzed a special class of dynamical maps represented by the generalized Pauli channels. Such evolution can be realized for $d$-level quantum systems when the corresponding Hilbert space $\mathbb{C}^d$ admits the maximal number ($N(d)=d+1$) of MUBs. We derived both necessary and sufficient conditions for vanishing negative flow of information which generalize well known conditions in the qubit case. Contrary to the qubit case ($d=2$) necessary condition is no longer sufficient. Interestingly, it turns out that among the set $\{\gamma_1(t),\ldots,\gamma_{d+1}(t)\}$  one may have $d-1$ $\gamma$s which are negative and  still conditions (\ref{P-d}) are satisfied. This paper clearly shows that even for relatively simple dynamical maps -- as generalized Pauli channels -- the precise characterization of Markovianity conditions based on the condition (\ref{BLP}) is highly involved.
%
%

It would be interesting to analyze multi-partite evolution defined by generalized Pauli channels in connection to evolution of quantum correlations.

\vspace{.2cm}

\section*{Acknowledgements} This paper was partially supported by the National Science Centre project 2015/17/B/ST2/02026.

\end{document}